# All-optical spin-selective control and shaping of frequency conversion pathways

Xiaoxi Xu[1*], Mai Tal[1,2], Danielle Ben-Haim[1], Tal Ellenbogen[1*]

1. *Department of Physical Electronics, School of Electrical & Computer Engineering, Tel-Aviv University, Tel-Aviv 6779801, Israel* and *Center for Light-Matter Interaction, Tel-Aviv University, Tel-Aviv 6779801, Israel*

2. *Department of Condensed Matter Physics, School of Physics and Astronomy, Tel-Aviv University, Tel-Aviv 6779801, Israel*

**Corresponding author e-mail: xuxiaoxi99@gmail.com; tellenbogen@tauex.tau.ac.il*

## Abstract

The interplay between optical spin and crystal symmetry provides a powerful mechanism to control light-matter interaction, enabling advances in imaging, holography, sensing, signal processing, and more. Here we reveal that this interplay can also be exploited as a programmable mechanism for selecting and shaping nonlinear frequency-conversion pathways. Using a z-cut lithium niobate thin film, we experimentally demonstrate spin-selective switching between complementary three-wave mixing pathways, where co- and cross-circular-polarization excitations activate sum- and difference-frequency generation, respectively. We further generalize this concept to establish a quantitative spin-to-frequency mapping, that encodes the pump spin composition into the generated frequencies, and enables the reverse retrieval of the pump spin state. Finally, we leverage this concept also to demonstrate mapping of spin patterns to a combination of frequency-dependent real space images, and Fourier-space diffraction patterns. These results pave the way for all-optical programable frequency-conversion and spatiotemporal nonlinear signal manipulation in free-space and integrated photonic systems.

## INTRODUCTION

The ability to control the characteristics of light signals is a fundamental requirement in modern photonics and serves as the basis for a wide range of applications, from optical communications and information processing to integrated photonic technologies(*1*, *2*). While optical modulation is traditionally achieved through electro-optic(*3*, *4*), thermo-optic(*5*), or acousto-optic(*6*) effects, increasing demands for speed and integration have stimulated growing interest in all-optical control schemes(*7*, *8*). In such schemes, the control process is usually mediated by the nonlinear interaction between optical fields, and thus achieving ultrafast response(*9*, *10*). Among the many nonlinear platforms, second-order $\left(\chi^{(2)}\right)$ nonlinear materials have attracted significant attention as they support efficient three-wave mixing processes, such as second harmonic, sum-frequency and difference-frequency generation (SFG, DFG)(*11*, *12*). These processes not only form the basis for coherent frequency conversion over the entire optical spectrum, but also provide a direct way to establish controllable coupling between the interacting optical fields(*9–16*). Extensive efforts have been devoted to develop schemes to enhance these nonlinear processes through controlled propagation, such as birefringent phase matching(*16*), modal phase matching(*17*), quasi-phase matching(*18*), and geometric phase matching(*19*, *20*). These approaches allow to selectively enhance a given nonlinear conversion pathway over all the other possibilities through coherent accumulation. However, their selectivity generally relies on propagation over a sufficient interaction length, making them less effective in nanoscale and compact systems.

Recent advances in nano-optics and metasurfaces have opened new avenues for controlling nonlinear optical processes(*21*, *22*). Designing artificial structures at the subwavelength scale enables to tailor local electromagnetic fields, engineer resonant mode coupling, and manipulate the symmetry of light-matter interactions(*23–26*). Using geometric anisotropy and artificial symmetry breaking, researchers have realized a variety of nonlinear functionalities, including polarization-selective second-harmonic generation, nonlinear wavefront shaping, nonlinear holography, and spatial routing of nonlinear outputs(*24*, *27–30*). Meanwhile, advances in two-dimensional materials and thin film crystalline platforms have further broadened the scope of nonlinear photonics, enabling diverse nonlinear functionalities and integrated photonic applications(*31–34*). While these developments have greatly enriched the available tools for nonlinear photonics, achieving simple route for selective control over desired nonlinear frequency-conversion pathways remains a challenge.

The spin state of light, corresponding to left- or right-hand circular polarization (LCP or RCP), constitutes a robust binary degree of freedom that can be used for optical information encoding and modulation(*35*, *36*). In recent years, spin (polarization)-controlled nonlinear optics has received increasing amount of attention(*37–40*). Researchers have explored various strategies, including plasmonic metasurfaces(*22*, *27*), dielectric nanostructures with customized symmetries(*29*), and nonlinear photonic crystals(*41*), achieving polarization-dependent harmonic generation, directional control, and nonlinear routing functionalities(*42–44*). Here, we demonstrate, for the first time to our knowledge, all-optical spin-selective activation of nonlinear frequency conversion pathways in an unstructured thin film lithium niobate ($LiNbO_3$). Through the coupling between optical spin and the intrinsic $\chi^{(2)}$ tensor of z-cut $LiNbO_3$, we show that the relative spin configuration of the pump and signal deterministically selects between SFG and DFG pathways. Extending the discrete spin-selection rules to arbitrary polarization states establishes a continuous spin-to-frequency mapping, in which the spin composition of the input fields is encoded in the frequency-resolved SFG and DFG intensities. This forward mapping further enables retrieval of the pump spin composition from the measured nonlinear outputs. Finally, by engineering the spin composition and spatial distribution of the pump, we demonstrate that it allows also spatial encoding of frequency-converted nonlinear signals in real space and momentum space opening the door to new optical functionalities.

## RESULTS

### Spin selection of sum- and difference-frequency generation in Lithium Niobate

For plane wave propagating along the crystal z-axis ($E_z = 0$), the second-order nonlinear response of a z-cut $LiNbO_3$ crystal (point group 3m, crystal structure shown in Fig. 1A) is governed by the transverse nonlinear susceptibility tensor. The corresponding nonlinear polarization for the SFG process is given by

$$\begin{bmatrix} P_x(\omega_i) \\ P_y(\omega_i) \\ P_z(\omega_i) \end{bmatrix} = 2\epsilon_0 \begin{pmatrix} 0 & 0 & 0 & 0 & d_{31} & -d_{22} \\ -d_{22} & d_{22} & 0 & d_{31} & 0 & 0 \\ d_{31} & d_{31} & d_{33} & 0 & 0 & 0 \end{pmatrix} \begin{pmatrix} E_x(\omega_p)E_x(\omega_s) \\ E_y(\omega_p)E_y(\omega_s) \\ 0 \\ 0 \\ 0 \\ E_x(\omega_p)E_y(\omega_s) + E_y(\omega_p)E_x(\omega_s) \end{pmatrix}, (1)$$

where $\omega_p$, $\omega_s$, and $\omega_i = \omega_p + \omega_s$ denote the pump, signal, and generated SFG frequencies, respectively, and $\epsilon_0$ is the vacuum permittivity. The nonlinear coefficients(*45*) are $\mathrm{d}_{22} = 2.1\ \mathrm{pm/V}$, $\mathrm{d}_{31} = 4.6\ \mathrm{pm/V}$, $\mathrm{d}_{33} = 25.2\ \mathrm{pm/V}$. For DFG at $\omega_i = \omega_p - \omega_s$, the corresponding nonlinear polarization takes an analogous form, with the signal field transformed into its complex conjugate, $E^*(\omega_s)$.

To reveal the symmetry-imposed spin selection rules, we write the interacting fields in the circular polarization basis

$$E_\sigma(\omega) = \frac{1}{\sqrt{2}}\left(E_x(\omega) + i\sigma E_y(\omega)\right), \tag{2}$$

where $\sigma = \pm 1$ denotes the optical spin state (left- or right- handed circular polarization). Substituting Eq. (2) into Eq. (1) yields spin-controlled selection rules for the nonlinear frequency-conversion pathway. When the pump and signal possess identical spin states $\sigma_p = \sigma_s$, only the SFG pathway is activated, while the DFG completely vanishes. In this case, the nonlinear polarizations associated with the SFG and DFG processes can be written as

$$P_{-\sigma}^{\mathrm{SFG}} = -4\sigma\epsilon_0 d_{22} i E_\sigma(\omega_p) E_\sigma(\omega_s),$$
$$P^{\mathrm{DFG}} = 0. \tag{3}$$

In contrast, when the pump and signal carry opposite spin states $\sigma_p = -\sigma_s$, the SFG pathway is suppressed, and the DFG pathway becomes activated. In this case, the corresponding nonlinear polarizations take the form

$$P^{\mathrm{SFG}} = 0,$$
$$P_\sigma^{\mathrm{DFG}} = 4\sigma\epsilon_0 d_{22} i E_{-\sigma}(\omega_p) E_\sigma^*(\omega_s). \tag{4}$$

Thus, co-circular polarization excitation selectively activates SFG, whereas cross-circular polarization excitation selectively activates DFG. This is not merely a spin-dependent variation in conversion efficiency, as in the case of conventional phase matching. Instead, the relative input-spin configuration determines which of the two nonlinear frequency-conversion pathways is symmetry allowed. This selection rule can also be understood from spin angular momentum conservation in the nonlinear interaction, enforced by the threefold rotational symmetry ($C_3$) of the z-cut $LiNbO_3$ crystal. In such a system, the spins of the interacting waves must satisfy the symmetry-imposed relations

$$\sigma_p + \sigma_s - \sigma_{SFG} = 3m, \tag{5}$$

for SFG, and

$$\sigma_p - \sigma_s - \sigma_{DFG} = 3m, \quad (6)$$

for DFG, where $m$ is an integer associated with the threefold symmetry of the lattice. Because each optical spin takes the value $\pm 1$, the left-hand sides of Eqs. (5) and (6) can take the values $-3, -1, +1$ or $+3$. Only the values $\pm 3$ satisfy the $C_3$ symmetry constraint, corresponding to $m = \pm 1$.

The symmetry-imposed selection rules establish a four-state spin-selective mapping between the input spin states $(\sigma_p, \sigma_s)$ and output frequency states (SFG, DFG), as summarized in Fig.1B. Co-circular input spin configurations $(\sigma_p, \sigma_s) = \pm(1, 1)$ activate SFG with opposite output spins, whereas cross-circular configurations $(\sigma_p, \sigma_s) = \pm(1, -1)$ activate DFG with opposite spins. This opens the door to new applications of programmable frequency conversion as we demonstrate in what follows.

**Experimental demonstration of spin-selective pathway switching**

We start with one of the input spin states fixed, e.g. taking $\sigma_s = +1$, and change $\sigma_p$ between $+1$ and $-1$. In this case the nonlinear response is governed solely by the spin state of the pump. This provides a direct and simple route for all-optical modulation of the SFG and DFG nonlinear interaction pathways as illustrated in Fig. 1C. To experimentally verify the proposed symmetry-controlled nonlinear interaction, we employ a 500-nm-thick z-cut $LiNbO_3$ thin film on a silica substrate. A pump at 438 nm and a signal at 1600 nm drive the three-wave mixing process. The specific wavelengths of the pump and signal are not unique and were chosen to fit our experimental system and detection capabilities. The subwavelength thickness of the $LiNbO_3$ film effectively relaxes phase-matching constraints, thereby enabling direct observation of the symmetry-imposed spin-selection rules(*16*). The experimental configuration is illustrated in Fig. 1D, where the polarization states of the pump and signal are independently controlled to prepare different spin combinations. The generated nonlinear spectra are then measured under these conditions to identify the corresponding nonlinear interaction pathways.

Figure 1E presents the measured spectrum for pump polarization state $\sigma_p = +1$. In this case only the SFG signal is observed at 344 nm, while no DFG signal detected. Reversing the pump spin to the opposite handedness ($\sigma_p = -1$) switches the nonlinear frequency-conversion pathway from SFG to DFG, with the SFG signal fully suppressed and a distinct DFG peak appearing at 603 nm, as displayed in Fig. 1F. The measured spectra are in excellent agreement with the symmetry-imposed selection rules derived in last section, providing direct experimental validation of the spin-selective activation of nonlinear frequency-conversion pathways concept.

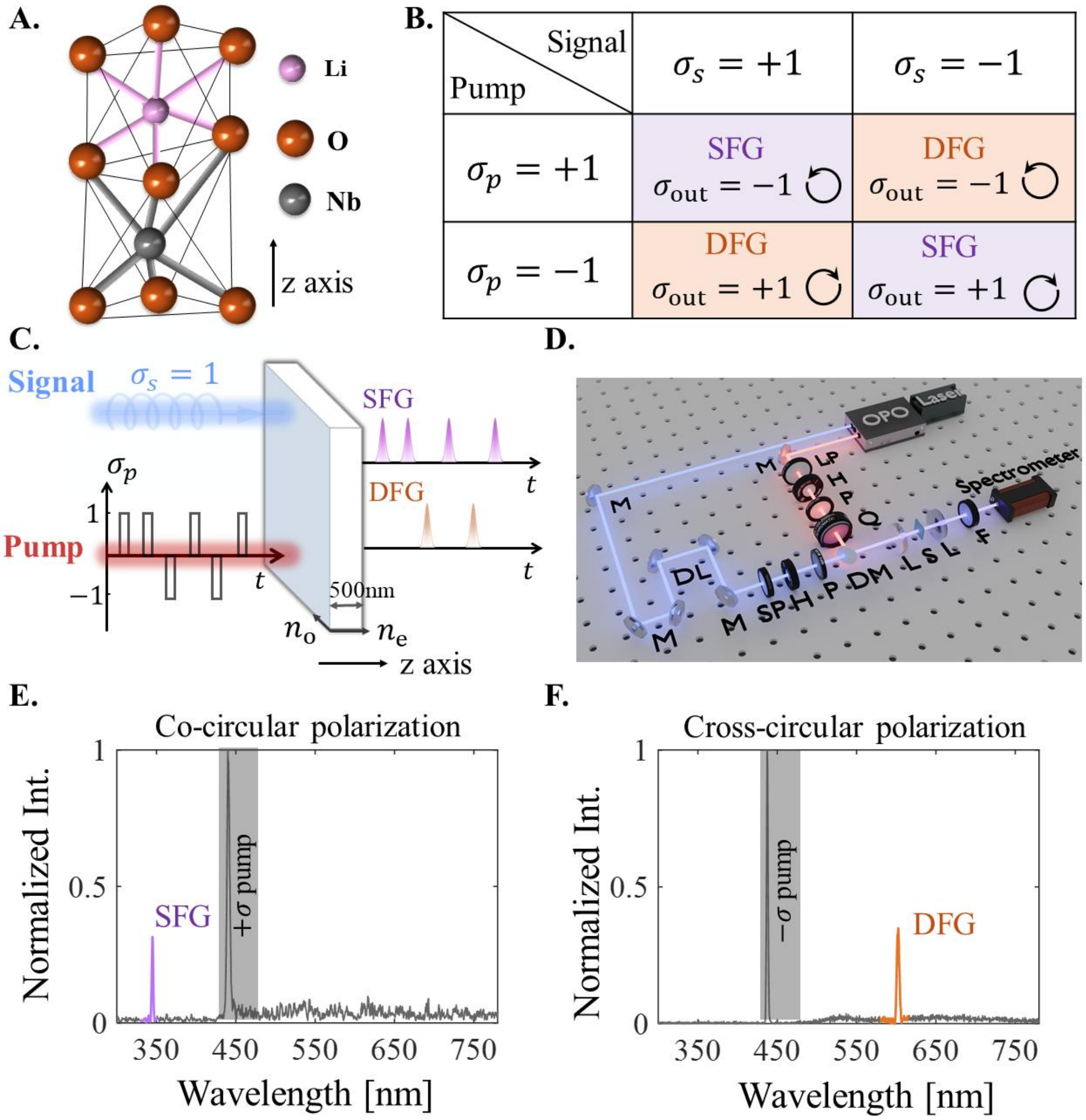

**Figure 1. Concept of spin-selective activation of nonlinear frequency-conversion pathways in z-cut $LiNbO_3$ and experimental demonstration**. **(A)** The crystal structure of z-cut lithium niobate with threefold rotational symmetry. **(B)** Spin-selective mapping between the input spin configuration and the generated nonlinear state in z-cut $LiNbO_3$. **(C)** Schematic illustration of the spin-controlled dynamic modulation of SFG and DFG nonlinear frequency-conversion pathways. **(D)** Experimental setup. OPO, optical parametric oscillator; M, mirror; DL, delay stage; LP, long-pass filter; SP, short-pass filter; H, half-waveplate; P, linear polarizer; Q, quarter-waveplate; DM, dichroic mirror; L, lens; S, sample; F, filter. **(E)** Experimentally measured nonlinear spectrum under co-circularly polarized excitation $(\sigma_p, \sigma_s) = (+1, +1)$, showing selective generation of the SFG signal at 344 nm. **(F)** Experimentally measured nonlinear spectrum after switching the pump spin to achieve cross-circularly polarized excitation $(\sigma_p, \sigma_s) = (-1, +1)$, showing selective generation of the DFG signal at 603 nm. The spectra are normalized to their respective maximum intensities. The shaded regions denote the residual pump spectrum.

**Continuous spin-to-frequency mapping and spin retrieval**

To examine the continuous evolution of the spin-selective nonlinear response, we gradually tune the pump polarization from dominant $\sigma^-$ state to dominant $\sigma^+$ state while keeping the signal fixed at $\sigma_s = +1$. An arbitrarily polarized pump can be expressed as a coherent superposition of two optical spin states $\sigma = +1$ and $\sigma = -1$. Their respective fractions are denoted by $f_{\sigma^+}$ and $f_{\sigma^-}$, satisfying $f_{\sigma^+} + f_{\sigma^-} = 1$. Representative nonlinear spectra for three characteristic pump states are shown in Fig. 2A. The DFG and SFG intensities evolve as the pump spin composition is varied. When the pump was predominantly $\sigma_p^-$ polarized (inset (a)), the DFG pathway dominates, whereas the SFG signal is strongly suppressed. As the pump approaches an equal superposition of the two spin components (inset (b)), it becomes nearly linearly polarized, and both SFG and DFG are simultaneously generated. When the pump was predominantly $\sigma_p^+$ polarized (inset (c)), the SFG pathway becomes dominant and the DFG output is suppressed. The continuous evolution is quantified in Fig. 2B. The SFG intensity increases linearly with the fraction of the $\sigma = +1$ component, $f_{\sigma^+}$, while the DFG output exhibits a corresponding linear dependence on $f_{\sigma^-}$. The agreement between the experimental data and the linear fits confirms the quantitative spin-to-frequency mapping dictated by the symmetry-imposed selection rules.

This mapping can also be inverted as a metrology means to retrieve the spin composition of an unknown pump. More generally, the spin-selection rules map the circular components of both input fields onto four frequency- and spin-resolved nonlinear output channels. In principle, measuring these four channels allows the spin-component weights of both input fields to be determined, as described in Supplementary Note 1. In the present experiment, the signal is fixed at $\sigma_s = +1$, reducing the general four-channel problem to two measurable output channels: SFG selectively probes the $\sigma^+$ component of the pump, whereas DFG selectively probes its $\sigma^-$ component. Because the two nonlinear pathways may have different conversion efficiencies and detection responses, their measured intensities are first independently calibrated. The pump spin-component weights are then obtained from

$$f_{p,\sigma^+} = \frac{\tilde{I}_{\mathrm{SFG}}}{\tilde{I}_{\mathrm{SFG}} + \tilde{I}_{DFG}}, f_{p,\sigma^-} = \frac{\tilde{I}_{\mathrm{DFG}}}{\tilde{I}_{\mathrm{SFG}} + \tilde{I}_{\mathrm{DFG}}}, \tag{7}$$

where $\tilde{I}_{\mathrm{SFG}}$ and $\tilde{I}_{\mathrm{DFG}}$ are the calibrated measured intensities of SFG and DFG, respectively. The detailed derivation analysis can be found in Supplementary Note 1. Fig. 2C visualizes this reverse mapping in the calibrated SFG-DFG intensity space. The upper-left region corresponds to a $\sigma^+$ dominant pump, whereas the lower-right region corresponds to a $\sigma^-$ dominant pump. The diagonal defined by $\tilde{I}_{\mathrm{SFG}} = \tilde{I}_{\mathrm{DFG}}$ corresponds to a vanishing normalized third Stokes parameter, $\frac{S_{3,p}}{S_{0,p}} = 0$, and hence to a linearly polarized pump. The three representative states in Fig. 2A are marked on the map, showing how the nonlinear outputs provide a direct spin-resolved readout of the pump. Specifically, $\frac{S_{3,p}}{S_{0,p}} = 1$ and $\frac{S_{3,p}}{S_{0,p}} = -1$ correspond to pure $\sigma^+$ (RCP) and pure $\sigma^-$ (LCP) pump, respectively. Intermediate values, $0 < \left|\frac{S_{3,p}}{S_{0,p}}\right| < 1$ indicate unequal circular-component weights and correspond to elliptically polarized pumps.

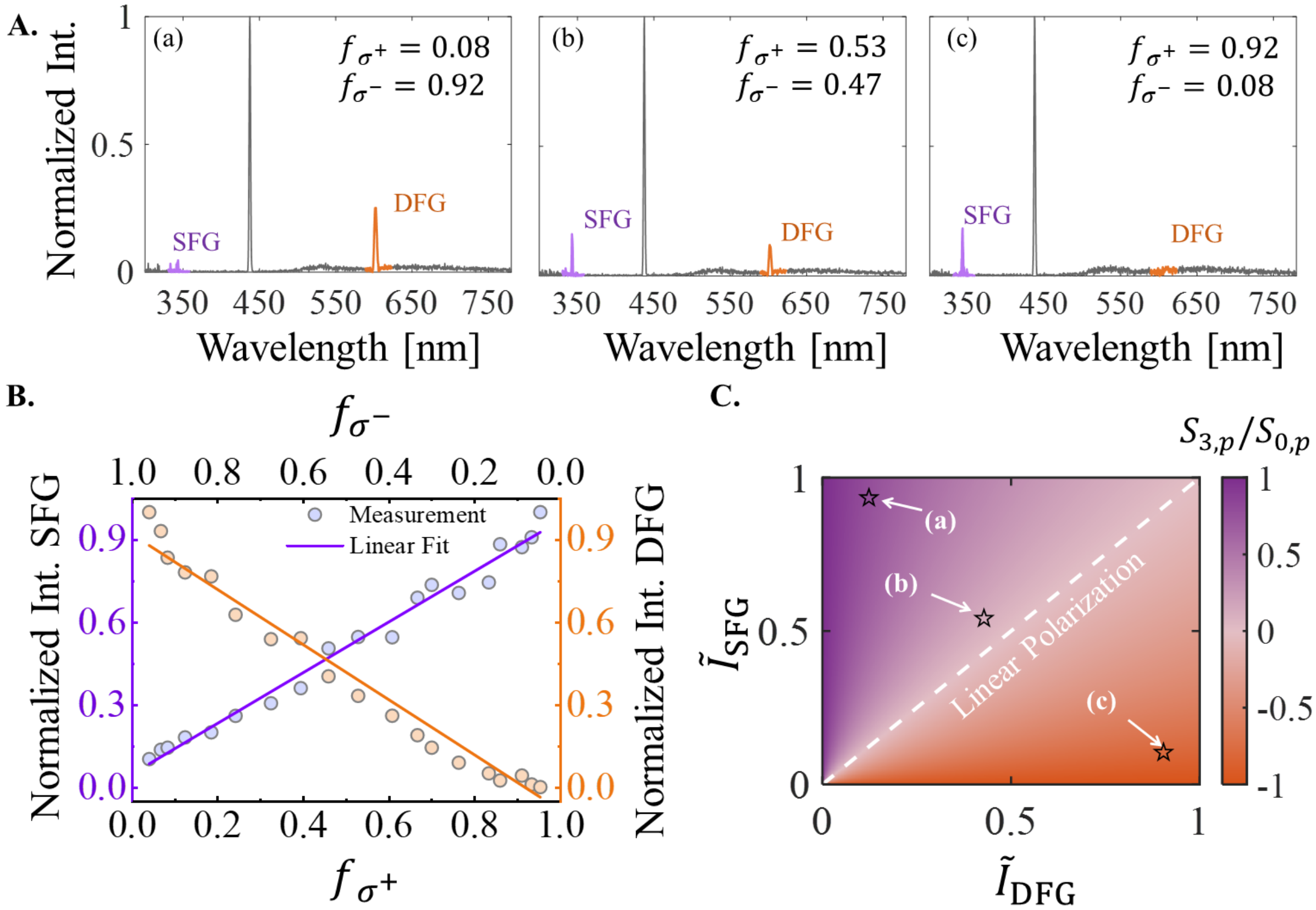

**Figure 2. Continuous spin-to-frequency mapping and reverse retrieval of pump spin composition.** (**A**) Representative nonlinear spectra for three pump polarization states with a fixed $\sigma_s = +1$ signal: $\sigma_p^-$ dominant elliptical state, $(f_{\sigma^+}, f_{\sigma^-}) = (0.08, 0.92)$, nearly linear polarization state $(f_{\sigma^+}, f_{\sigma^-}) = (0.53, 0.47)$, and $\sigma_p^+$ dominant elliptical polarization $(f_{\sigma^+}, f_{\sigma^-}) = (0.92, 0.08)$. The spectra are normalized to their respective maximum intensities. (**B**) Quantitative spin-to-frequency mapping. The independently normalized SFG and DFG intensities are plotted against the pump circular-component fractions $f_{\sigma^+}$ and $f_{\sigma^-} = 1 - f_{\sigma^+}$, respectively. The circles show the experimental measurements, and the solid lines are linear fits. (**C**) Reverse retrieval of the pump spin composition from the calibrated nonlinear outputs. The color map shows the retrieved normalized third Stokes parameter $\frac{S_{3,p}}{S_{0,p}} = f_{p,\sigma^+} - f_{p,\sigma^-}$ as a function of $\tilde{I}_{\mathrm{SFG}}$ and $\tilde{I}_{\mathrm{DFG}}$. The dashed diagonal corresponds to equal circular-component weights, $\frac{S_{3,p}}{S_{0,p}} = 0$, while the marked

points indicate the three representative states shown in **(A)**. The derivation of the retrieval relation and its four-channel generalization are provided in Supplementary Note 1.

### Spin to frequency nonlinear imaging

The demonstrated dependence of the nonlinear response on the pump spin composition naturally enables temporal as well as spatial control of frequency-conversion. By engineering a spatially varying spin distribution, different nonlinear interaction pathways can be selectively activated at different spatial locations. To realize this concept, a binary phase pattern is encoded onto a spatial light modulator (SLM), and transformed into a spatially structured spin distribution of the pump, by using a polarizer and a waveplate, as shown in Fig. 3A. The signal is fixed in the $\sigma^+$ state throughout the experiment. According to the spin-controlled selection rules in the z-cut $LiNbO_3$ film, the pump regions with $\sigma^+$ polarization activate the SFG pathway, whereas the regions carrying $\sigma^-$ polarization activate the DFG. Consequently, the spatial spin pattern encoded on the pump and imaged on the crystal is directly mapped onto the nonlinear spatial frequency conversion output.

Figures 3B(a, b) display the experimentally measured DFG intensity distributions for two representative spatial masks. In Fig. B(a), the letter "X" is encoded in the $\sigma^-$ component of the pump. Therefore, the "X" region selectively activates the DFG pathway, producing a bright "X" pattern in the DFG image. In Fig. 3B(b), a complementary mask is used, where the "X" region carries the $\sigma^+$component while the surrounding region carries $\sigma^-$. As a result, DFG is suppressed in the "X" region and mainly generated in the surrounding background, leading to the inverted DFG contrast. The corresponding normalized SFG distributions, shown in Figs. 3B(c, d), are reconstructed from the measured DFG images using the complementary spin relations from Eq. (7), and $f_{\sigma^+} + f_{\sigma^-} = 1$. We used the reconstruction due to limited detection sensitivity of the imaging system in the SFG wavelength of 344 nm. These results confirm that spatial information encoded in optical spin can be routed into distinct nonlinear interaction pathways and generate frequency dependent images.

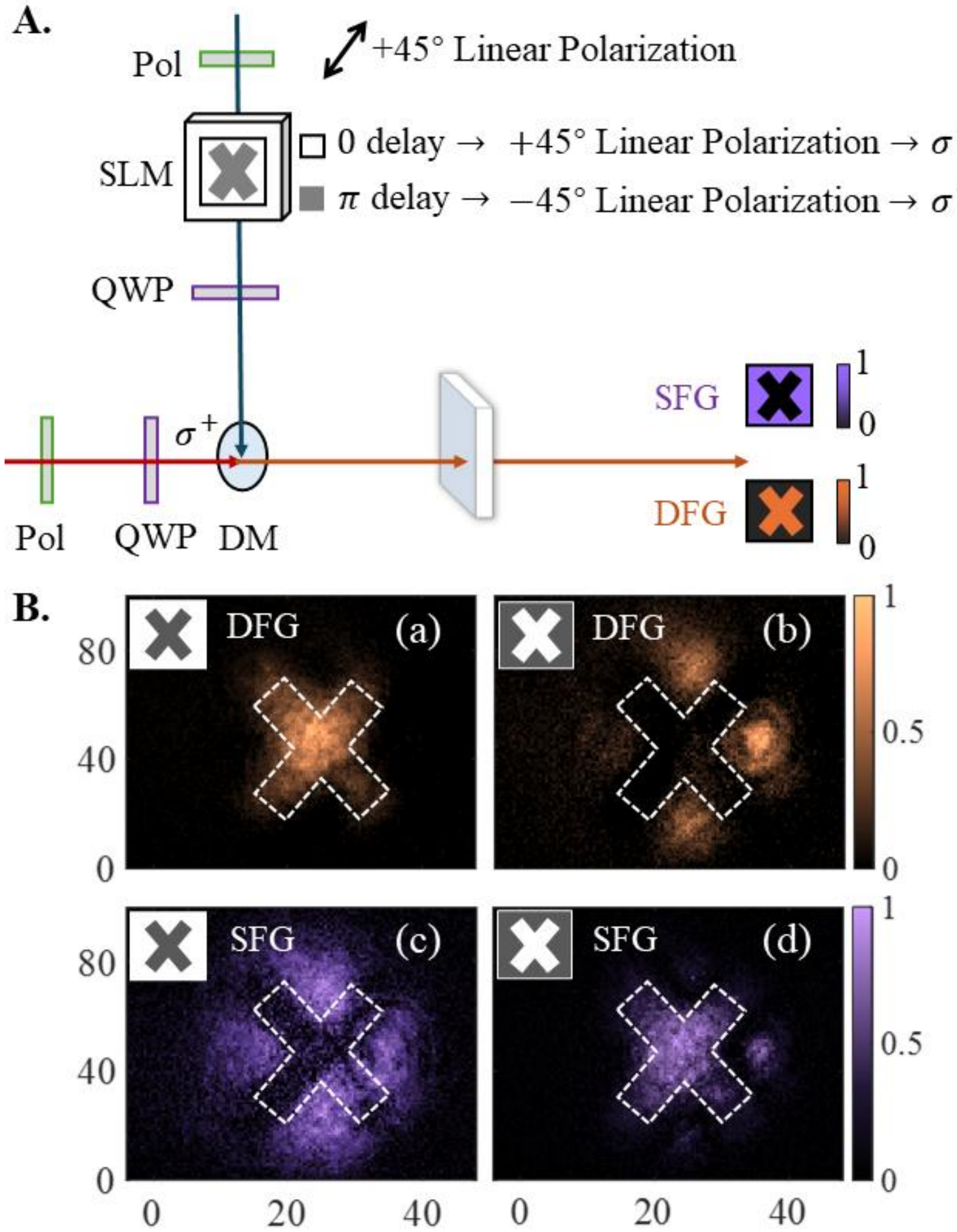

**Figure 3. Spin-controlled spatial modulation of nonlinear interaction pathways. (A)** Experimental implementation of spatial spin encoding using a spatial light modulator (SLM). **(B)** Experimentally captured DFG images generated from the encoded spin patterns. Corresponding SFG images reconstructed from the measured DFG patterns using the spin-selection rules. The symbols in the upper-left corners indicate the spatial patterns encoded on the SLM, and the dashed outlines mark the corresponding target profiles.

### Spin-selective nonlinear frequency diffraction in momentum space

We next show that the spin to frequency mapping can also be used to selectively and independently control the frequency diffraction in momentum space as illustrated in Fig. 4A. To this end, we consider a periodically structured modulation of the pump beam by the SLM consisting of two alternating polarization domains. In this case, the pump field is written as

$$\boldsymbol{E}_p(x, y, \omega_p) = \frac{E_{p,0} A_p(x, y)}{\sqrt{2}} [\boldsymbol{e}_{\sigma^+} + g(x)\boldsymbol{e}_{\sigma^-}], \tag{8}$$

where $A_p(x, y)$ is the spatial envelope of the pump and $g(x) = \pm 1$ is a 50%-duty-cycle binary periodic function with period $\Lambda$. Its definition is given in Supplementary Note 2. The two neighboring domains correspond to orthogonal linear polarization states, as shown in Fig. 4B. In the circular basis, the $\sigma^+$ component remains spatially in phase, whereas the $\sigma^-$ component acquires an alternating $0/\pi$ phase. The two spin components have identical local intensities, and their normalized spin fractions satisfy $f_{\sigma^+} = f_{\sigma^-} = 0.5$. Thus, the periodic structure is encoded in their relative spatial phase rather than their local spin composition. The complete circular-basis decomposition is provided in Supplementary Note 2. The signal is assumed to be spatially uniform with $\sigma_s = +1$consistent with the experimental configuration. The corresponding results for the reversed signal spin, $\sigma_s = -1$ , are presented in Supplementary Note 4 and Supplementary Fig. S2.

According to the established selection rules, the spatially uniform $\sigma^+$ pump component is mapped onto the SFG pathway, whereas the binary-phase $\sigma^-$ component is mapped onto the DFG pathway. Consequently, the two nonlinear polarizations excited in the thin film have identical intensity but different spatial phase, as shown in Fig. 4C. The SFG nonlinear polarization is spatially in-phase, whereas the DFG nonlinear polarization forms a binary $0/\pi$ nonlinear phase grating. The momentum space field at each generated frequency is proportional to the transverse Fourier transform of the corresponding nonlinear polarization

$$\widetilde{\boldsymbol{E}}_\gamma\left(k_x, k_y\right) \propto \mathcal{F}\left\{P_\gamma^{(2)}(x, y)\right\}, \qquad \gamma = \mathrm{SFG}, \mathrm{DFG} \tag{9}$$

For a binary grating with period $\Lambda$, diffraction peaks occur at

$$k_{x,n} = nG, \qquad G = \frac{2\pi}{\Lambda}, \qquad n = \pm 1, \pm 3, \pm 5, \cdots \tag{10}$$

As shown in Fig. 4D, the spatially uniform phase of the SFG nonlinear polarization produces an on-axis zeroth-order field. By contrast, the binary DFG phase grating completely eliminates the zeroth order and distributes the DFG among the odd diffraction orders $n = \pm 1, \pm 3, \pm 5, \cdots$ in agreement with Eq. (10) with the symmetric $n = \pm 1$ orders carrying the largest fraction. The predicted dependence of the first-order transverse momentum on the polarization-grating period is verified numerically in Supplementary Fig. S1(B). The dependence of the diffraction efficiencies on the binary phase difference, including the ideal combined first-order efficiency of $\frac{8}{\pi^2}$, is derived and analyzed in Supplementary Notes 2-3. These results extend the demonstrated spin-to-frequency mapping from spatially resolved imaging to momentum-space control.

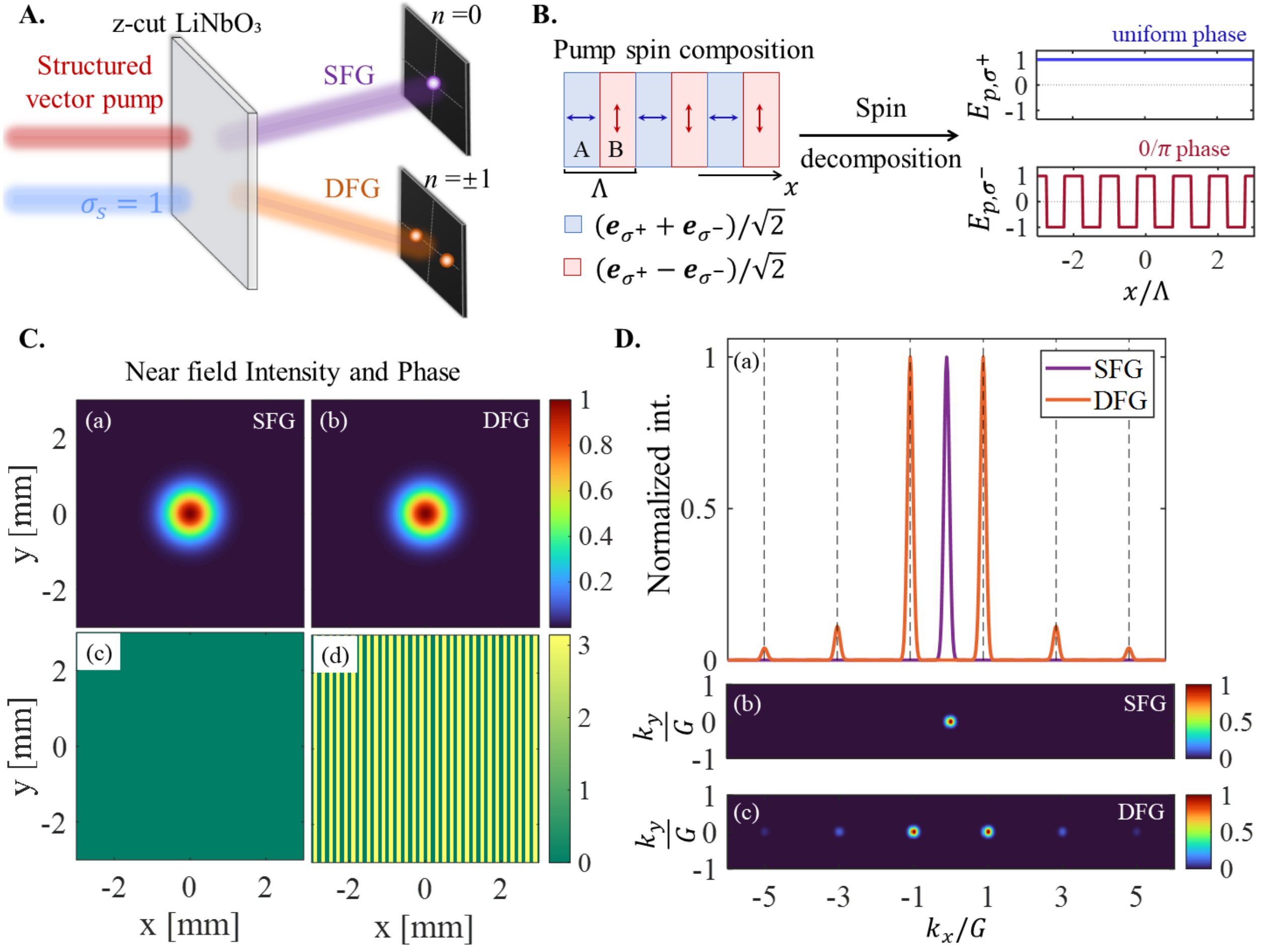

**Figure 4. Demonstration of spin-selective nonlinear phase-to-momentum mapping**. **(A)** Schematic of the nonlinear frequency-momentum mapping for the signal spin state $\sigma_s = +1$. The SFG output is concentrated in the zeroth order, whereas the dominant DFG output appears in the symmetric $n = \pm 1$ orders. **(B)** Periodically, structured vector pump consisting of alternating $\frac{e_{\sigma^+}+e_{\sigma^-}}{\sqrt{2}}$ (domain A) and $\frac{e_{\sigma^+}-e_{\sigma^-}}{\sqrt{2}}$ (domain B) polarization states with period Λ, together with its circular-basis decomposition. **(C)** Simulated near-field intensity and phase distributions of the SFG and DFG nonlinear polarizations, showing identical envelopes in (a, b). and showing a uniform SFG phase and a binary 0/π DFG phase grating in (c, d). **(D)** Simulated diffraction of SFG and DFG in momentum space. (a) Fourier-plane line profiles along the modulation direction. Dashed lines indicate the predicted odd diffraction orders, $k_{x,n} = \frac{2\pi n}{\Lambda}$. (b, c) Fourier-plane intensity distributions of the SFG and DFG.

## DISCUSSION

In conclusion, we have theoretically and experimentally demonstrated deterministic, spin-selective control, of nonlinear frequency-conversion pathways in a z-cut $LiNbO_3$ thin film, and translated this control into several practical implementations. By exploiting intrinsic crystal symmetry, we show that the relative spin configuration of the interacting waves selectively activates the SFG and DFG pathways. This provides a distinct mechanism for allowing only a specific frequency conversion pathway, without tailoring the material structure or relying on conventional phase matching. Therefore, it does not rely on the interaction length and can be used also in nanoscale and compact systems. We also establish a quantitative spin-to-frequency mapping by extending the spin-selection rules from discrete spin configurations to arbitrary input states. This provides a new method for retrieving the input spin composition through spectral measurements. We further extend the concept to demonstrate nonlinear spin to frequency selective imaging and diffraction of SFG and DFG. This lays the foundations to new types of spin-selective nonlinear interactions of vector beams as well as new types of nonlinear holography. Our results therefore establish optical spin as a new degree of control for accessing and spatially programming nonlinear frequency-conversion pathways. This can be applied for nonlinear signal processing, reconfigurable frequency routing, spin-selective nonlinear holography, structured nonlinear wavefront generation, and all-optical information processing. The same concept may also be extended to other $\chi^{(2)}$ crystals, layered nonlinear materials, 2D materials, or nonlinear metasurfaces possessing various rotational symmetries.

## MATERIALS AND METHODS

A 500-nm-thick z-cut $LiNbO_3$ thin film with lateral dimensions of 6 mm × 6 mm, supported on a 0.5-mm-thick silica substrate, was used throughout this work. The optical source was a femtosecond optical parametric oscillator (Chameleon Compact OPO; pulse duration, 200 fs; repetition rate, 80 MHz) pumped by a Ti:Sapphire laser at 876 nm, providing two beams at 438 nm and 1600 nm that served as the pump and signal, respectively. In the signal arm, a long-pass filter (FELH1250) was used to remove residual 876 nm light, the power and polarization were controlled using a half-wave plate, a linear polarizer, and a quarter-wave plate. In the pump arm, a motorized delay stage was used to temporally overlap the pump and signal pulses, while a short-pass filter (FESH500) removed unwanted spectral components. The pump and signal beams were combined by a dichroic mirror (DMSP950R) at an incidence angle of $45^\circ$. After combination, the two beams were collinearly focused onto the sample at normal incidence using a lens with a focal length of 30 cm. The pump polarization was calibrated at the sample position by analyzing the transmitted polarization state after the dichroic mirror. The input wave plates were adjusted to compensate for its polarization-dependent phase retardation, ensuring the desired polarization state at the sample plane. After the nonlinear interaction, the generated frequency-converted signals were collected by a second lens with the same focal length. After residual pump and signal were removed by spectral filters, the nonlinear spectra were measured using a spectrometer (ANDOR Shamrock 303).

For the spatial modulation experiment, the pump was initially prepared in a $+45^{\circ}$ linearly polarized state with respect to the principal axes of the spatial light modulator (SLM). A binary phase mask consisting of 0 and $\pi$ phase delay was applied to the pump. After linear polarization, the encoded phase pattern generated two orthogonal linear polarization states ($+45^{\circ}$ and $-45^{\circ}$), which were subsequently converted into opposite spin states ($\sigma^{+}$ and $\sigma^{-}$) using a quarter-wave plate. This process generated a spatially varying spin distribution in the pump, while the signal was maintained in the $\sigma^{+}$ state throughout the measurements. The DFG intensity distribution was isolated by spectral filtering and recorded at the image plane.

## Acknowledgements

**Funding:** This research was supported by the European Research Council (ERC 3D NOAM 101044797). D.B.H acknowledges the doctoral scholarship given by the Council for Higher Education (CHE) of Israel. M.T. acknowledges the support of the Milner Foundation fellowship for PhD students. **Author contributions:** X.X. performed the theoretical analysis and experiments, analyzed the data, and wrote the original manuscript. M.T. and D.B.H provided guidance on the experiments, contributed to the discussion, and reviewed and edited the manuscript. T.E. conceived the original idea, supervised the project, guided the experimental design and development of the research, and reviewed and edited the manuscript. All authors discussed the results and approved the final version of the manuscript. **Competing interests:** The authors declare that they have no competing interests. **Data and materials availability:** All data needed to evaluate the conclusions in the paper are present in the paper and/or the Supplementary Materials.

# Supplementary Materials for

## All-optical spin-selective control and shaping of frequency conversion pathways

Xiaoxi Xu* et al

*Corresponding author. E-mail: xuxiaoxi99@gmail.com

### Note 1 Reverse retrieval of the input spin composition

The continuous evolution of the sum-frequency generation (SFG) and different-frequency generation (DFG) indicates that the spin-selective nonlinear mapping can be extended from the two circularly polarized eigenstates to an arbitrary input polarization. In the circular-polarization basis, an arbitrary pump field can be expressed as

$$\boldsymbol{E}_p(\omega_p) = E_{p,0}\left[\sqrt{f_{p,\sigma^+}}\boldsymbol{e}_{\sigma^+} + e^{i\varphi_p}\sqrt{f_{p,\sigma^-}}\boldsymbol{e}_{\sigma^-}\right]$$

$$\boldsymbol{E}_s(\omega_s) = E_{s,0}\left[\sqrt{f_{s,\sigma^+}}\boldsymbol{e}_{\sigma^+} + e^{i\varphi_s}\sqrt{f_{s,\sigma^-}}\boldsymbol{e}_{\sigma^-}\right], \tag{S1}$$

where $E_{p,0}, E_{s,0}$ are the amplitude of pump and signal, respectively, $\varphi_{p,s}$ is the relative phase between the two circularly polarized components, and $f_{p,\sigma^\pm}, f_{s,\sigma^\pm}$ are normalized circular-component intensity fractions. They satisfy

$$f_{p,\sigma^+} + f_{p,\sigma^-} = 1, \qquad f_{s,\sigma^+} + f_{s,\sigma^-} = 1. \tag{S2}$$

With the circular-basis convention used in the main text, the spin-selective nonlinear polarizations are

$$\begin{aligned}
P_-^{\text{SFG}} &= -4i\varepsilon_0 d_{22} E_{p,+}(\omega_p) E_{s,+}(\omega_s),\\
P_+^{\text{DFG}} &= 4i\varepsilon_0 d_{22} E_{p,-}(\omega_p) E_{s,+}^*(\omega_s),\\
P_+^{\text{SFG}} &= 4i\varepsilon_0 d_{22} E_{p,-}(\omega_p) E_{s,-}(\omega_s),\\
P_-^{\text{DFG}} &= -4i\varepsilon_0 d_{22} E_{p,+}(\omega_p) E_{s,-}^*(\omega_s).
\end{aligned} \tag{S3}$$

Because the generated intensity is proportional to the squared modulus of the corresponding nonlinear polarization, the measured intensities satisfy

$$I_-^{\text{SFG}} = C_-^{\text{SFG}} I_p I_s f_{p,\sigma^+} f_{s,\sigma^+}, \qquad I_+^{\text{SFG}} = C_+^{\text{SFG}} I_p I_s f_{p,\sigma^-} f_{s,\sigma^-},$$

$$I_-^{\text{DFG}} = C_-^{\text{DFG}} I_p I_s f_{p,\sigma^+} f_{s,\sigma^-}, \qquad I_+^{\text{DFG}} = C_+^{\text{DFG}} I_p I_s f_{p,\sigma^-} f_{s,\sigma^+}. \tag{S4}$$

Here, $I_p$ and $I_s$ denote the total intensities of the pump and signal fields at the sample, respectively. $C_{\pm}^{SFG}$ and $C_{\pm}^{DFG}$ are the channel-dependent proportionality coefficients.

**Generalization to the retrieval of two unknown input spin compositions**

For completeness, we first consider a generalized configuration in which both the output frequency and output spin are resolved. The calibrated channel signals are defined as

$$J_{++} = \frac{I_{-}^{\mathrm{SFG}}}{\eta_{-}^{\mathrm{SFG}}}, \qquad J_{--} = \frac{I_{+}^{\mathrm{SFG}}}{\eta_{+}^{\mathrm{SFG}}},$$

$$J_{+-} = \frac{I_{-}^{\mathrm{DFG}}}{\eta_{-}^{\mathrm{DFG}}}, \qquad J_{-+} = \frac{I_{+}^{\mathrm{DFG}}}{\eta_{+}^{\mathrm{DFG}}}. \tag{S5}$$

Ideally,

$$J_{\alpha\beta} = \mathcal{N} f_{p,\alpha} f_{s,\beta}, \qquad \alpha, \beta \in \{\sigma^+, \sigma^-\}, \tag{S6}$$

where $\mathcal{N}$ is proportional to $I_p I_s$. The normalized joint spin-response matrix is therefore

$$\mathbf{M} = \frac{1}{\mathcal{N}} \begin{pmatrix} J_{++} & J_{+-} \\ J_{-+} & J_{--} \end{pmatrix} = \begin{pmatrix} f_{p,\sigma^+} \\ f_{p,\sigma^-} \end{pmatrix} \begin{pmatrix} f_{s,\sigma^+} & f_{s,\sigma^-} \end{pmatrix}, \tag{S7}$$

with $\mathcal{N} = J_{++} + J_{+-} + J_{-+} + J_{--}$. The row and column marginals independently give all four circular-component fractions:

$$f_{p,\sigma^+} = \frac{J_{++} + J_{+-}}{\mathcal{N}}, \qquad f_{p,\sigma^-} = \frac{J_{-+} + J_{--}}{\mathcal{N}},$$

$$f_{s,\sigma^+} = \frac{J_{++} + J_{-+}}{\mathcal{N}}, \qquad f_{s,\sigma^-} = \frac{J_{+-} + J_{--}}{\mathcal{N}}. \tag{S8}$$

Only two of these four fractions are independent because each input separately obeys $f_{p,\sigma^+} + f_{p,\sigma^-} = 1, f_{s,\sigma^+} + f_{s,\sigma^-} = 1$. Their normalized third Stokes parameters are

$$\frac{S_{3,p}}{S_{0,p}} = f_{p,\sigma^+} - f_{p,\sigma^-} = \frac{J_{++} + J_{+-} - J_{-+} - J_{--}}{\mathcal{N}},$$

$$\frac{S_{3,s}}{S_{0,s}} = f_{s,\sigma^+} - f_{s,\sigma^-} = \frac{J_{++} + J_{-+} - J_{+-} - J_{--}}{\mathcal{N}}. \tag{S9}$$

**Reduction to the present experimental configuration**

In the present experiment, the signal is fixed at the known state $\sigma_s = 1$, for which

$$f_{s,\sigma^+} = 1, \qquad f_{s,\sigma^-} = 0. \tag{S10}$$

The general two-input mapping then reduces to a two-channel measurement: the $\sigma_p^+$ and $\sigma_p^-$ components of the unknown pump are selectively mapped onto the SFG and DFG outputs, respectively. After separately calibrating the two output channels,

$$\tilde{I}_{\mathrm{SFG}} = \frac{I_{\mathrm{SFG}}}{\eta_{\mathrm{SFG}}}, \qquad \tilde{I}_{\mathrm{DFG}} = \frac{I_{\mathrm{DFG}}}{\eta_{\mathrm{DFG}}}, \tag{S11}$$

the pump fractions are retrieved as

$$f_{p,\sigma^+} = \frac{\tilde{I}_{\mathrm{SFG}}}{\tilde{I}_{\mathrm{SFG}} + \tilde{I}_{\mathrm{DFG}}}, \qquad f_{p,\sigma^-} = \frac{\tilde{I}_{\mathrm{DFG}}}{\tilde{I}_{\mathrm{SFG}} + \tilde{I}_{\mathrm{DFG}}}. \tag{S12}$$

Consequently,

$$\frac{S_{3,p}}{S_{0,p}} = f_{p,\sigma^+} - f_{p,\sigma^-} = \frac{\tilde{I}_{\mathrm{SFG}} - \tilde{I}_{\mathrm{DFG}}}{\tilde{I}_{\mathrm{SFG}} + \tilde{I}_{\mathrm{DFG}}}, \tag{S13}$$

and the pump ellipticity angle is

$$\chi_p = \frac{1}{2}\arcsin\left(\frac{S_{3,p}}{S_{0,p}}\right) = \frac{1}{2}\arcsin\left(f_{p,\sigma^+} - f_{p,\sigma^-}\right). \tag{S14}$$

Thus, the calibrated SFG and DFG intensities provide a direct retrieval of the pump circular-component weights, normalized optical spin, and ellipticity.

## Note 2 Spin-selective nonlinear phase-to-momentum mapping

### Spin-selective nonlinear polarization

The periodically structured pump field is

$$\boldsymbol{E}_p(x, y, \omega_p) = \frac{E_{p,0} A_p(x, y)}{\sqrt{2}} [\boldsymbol{e}_{\sigma^+} + g(x) \boldsymbol{e}_{\sigma^-}], \tag{S15}$$

where $A_p(x, y)$ is the spatial envelope of the pump and $g(x)$ is a binary periodic function with period $\Lambda$ and a duty cycle of 50%:

$$g(x) = \begin{cases} +1, & n\Lambda \le x < n\Lambda + \frac{\Lambda}{2}, \\ -1, & n\Lambda + \frac{\Lambda}{2} \le x < (n+1)\Lambda, \end{cases} \tag{S16}$$

where $g(x + \Lambda) = g(x)$. The $\sigma^+$ component therefore remains spatially uniform in phase, whereas the $\sigma^-$ component acquires the binary phase modulation defined by $g(x)$. Because $|g(x)| = 1$, the modulation changes only the relative phase of the two spin components without altering their local populations. For the main-text configuration, the signal is spatially uniform with $\sigma_s = +1$,

$$\boldsymbol{E}_s(x,y,\omega_s) = E_{s,0}A_s(x,y)\boldsymbol{e}_{\sigma^+}, \tag{S17}$$

the SFG nonlinear polarization is

$$\boldsymbol{P}_{SFG}^{(2)}(x,y,\omega_p+\omega_s) = \epsilon_0\chi^{(2)}\boldsymbol{E}_{p,\sigma^+}\boldsymbol{E}_{s,\sigma^+}. \tag{S18}$$

Using the effective nonlinear coefficient, this becomes

$$\boldsymbol{P}_{SFG}^{(2)}(x,y) = \frac{\epsilon_0\chi_{eff}^{(2)}E_{p,0}E_{s,0}}{\sqrt{2}}A_p(x,y)A_s(x,y)\boldsymbol{e}_{SFG} \tag{S19}$$

Similarly, the DFG nonlinear polarization is

$$\boldsymbol{P}_{DFG}^{(2)}(x,y,\omega_p-\omega_s) = \epsilon_0\chi^{(2)}\boldsymbol{E}_{p,\sigma^-}\boldsymbol{E}_{s,\sigma^+}^{*}, \tag{S20}$$

which gives

$$\boldsymbol{P}_{DFG}^{(2)}(x,y) = \frac{\epsilon_0\chi_{eff}^{(2)}E_{p,0}E_{s,0}^{*}}{\sqrt{2}}A_p(x,y)A_s(x,y)g(x)\boldsymbol{e}_{DFG}. \tag{S21}$$

Because the pump and signal envelopes used in the numerical model are real-valued Gaussian functions, we define $A_{eff}(x,y) = A_p(x,y)A_s(x,y)$. For a general complex signal envelope, the DFG interaction contains $A_p(x,y)A_s^{*}(x,y)$. The spatial dependences can therefore be summarized as

$$P_{SFG}^{(2)}(x,y) \propto A_{eff}(x,y), \tag{S22}$$

and

$$P_{DFG}^{(2)}(x,y) \propto A_{eff}(x,y)g(x). \tag{S23}$$

Consequently, the SFG and DFG nonlinear polarizations have identical near-field intensity envelopes but different spatial phases. The SFG nonlinear polarization is spatially in phase, whereas the DFG nonlinear polarization carries a periodic binary $0/\pi$ phase modulation.

**Phase-dependent diffraction efficiency**

To examine the redistribution of the DFG power as a function of the phase difference between adjacent domains, we generalize the binary phase function as

$$t_{\Delta\phi}(x) = \begin{cases} 1, & 0 \le x < \dfrac{\Lambda}{2}, \\ e^{i\Delta\phi}, & \dfrac{\Lambda}{2} \le x < \Lambda, \end{cases} \qquad t_{\Delta\phi}(x+\Lambda) = t_{\Delta\phi}(x), \tag{S24}$$

The corresponding Fourier coefficients are

$$c_n(\Delta\phi) = \frac{1}{\Lambda}\int_0^{\Lambda} t_{\Delta\phi}(x)\exp\left(-i\frac{2\pi nx}{\Lambda}\right)dx. \quad \text{(S25)}$$

giving the zeroth-order diffraction efficiency

$$\eta_0 = |c_0|^2 = \cos^2\left(\frac{\Delta\phi}{2}\right). \quad \text{(S26)}$$

The diffraction efficiency of each odd order is

$$\eta_n(\Delta\phi) = |c_n|^2 = \frac{4}{\pi^2 n^2}\sin^2\left(\frac{\Delta\phi}{2}\right), \qquad n = \pm1, \pm3, \pm5, \cdots \quad \text{(S27)}$$

At $\Delta\phi = \pi$, the zeroth order ideally vanishes, whereas the combined $n \pm 1$ orders carry $\eta_{\pm1}^{\text{combined}}(\pi) = \frac{8}{\pi^2} \approx 0.811$. The remaining power is distributed among the higher odd diffraction orders. The numerical results agree with these analytical expressions, as shown in Supplementary Fig. S1 (A).

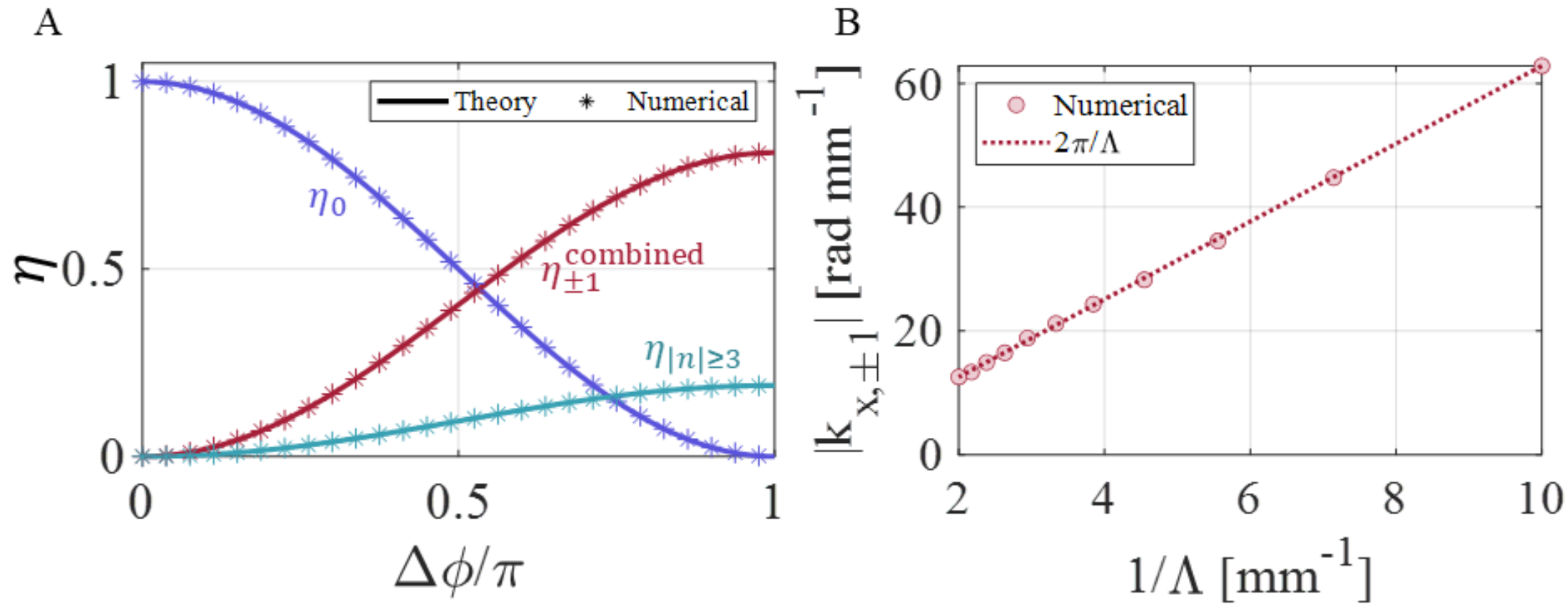

Figure S1 **(A)** Phase-dependent redistribution of the DFG diffraction. Numerically calculated zeroth-order, combined first-order and high-order diffraction efficiencies as functions of the phase difference $\Delta\phi$ between adjacent polarization domains. Solid curves and asterisk markers represent analytical and numerical results, respectively. Here, $\eta_{|n\geq3|} = \sum_{|n\geq3|, n\ odd}\eta_n = (1 - 8/\pi^2)\sin^2(\Delta\phi/2)$. **(B)** Control of the DFG transverse momentum by the polarization-grating period. Numerically extracted first-order transverse momentum $|k_{x,\pm1}|$ as a function of $\frac{1}{\Lambda}$. The dotted line represents the analytical relation $|k_{x,\pm1}| = \frac{2\pi}{\Lambda}$.

### Control of the nonlinear transverse momentum

For a binary grating with period $\Lambda$, the DFG diffraction peaks occur at

$$k_{x,n} = nG, \qquad G = \frac{2\pi}{\Lambda}, \qquad n = \pm1, \pm3, \pm5, \cdots \quad \text{(S28)}$$

In particular, the first-order transverse momenta satisfy $|k_{x,\pm1}| = \frac{2\pi}{\Lambda}$. Therefore, the transverse momentum is continuously tunable through the spatial period of the structured pump. To verify that the DFG transverse momentum is determined by the spatial period of the structured pump, we varied the polarization-grating period Λ and numerically extracted the position of the positive first-order diffraction peak. As shown in Fig. S1(B), the extracted transverse momentum increases linearly with $\frac{1}{\Lambda}$ and follows $|k_{x,\pm 1}| = \frac{2\pi}{\Lambda}$.

## Note 3 Finite-beam effects and numerical methods

For a finite Gaussian interaction envelope, the Fourier transform of the DFG nonlinear polarization is

$$\tilde{P}_{DFG}^{(2)}(k_x, k_y) \propto \tilde{A}_{eff}(k_x, k_y) * \tilde{g}(k_x), \tag{S29}$$

where $*$ denotes convolution. Using the Fourier coefficients of $g(x)$, the DFG field can be expressed as

$$\tilde{P}_{DFG}^{(2)}(k_x, k_y) \propto \sum_{\substack{n=-\infty \\ n\ odd}}^{+\infty} c_n \tilde{A}_{eff}\left(k_x - \frac{2\pi n}{\Lambda}, k_y\right). \tag{S30}$$

Thus, every diffraction order has a finite width determined by the Fourier transform of the pump-signal overlap envelope. The numerical model treated the 500-nm-thick z-cut $LiNbO_3$ film as a nonlinear polarization thin film. The pump and signal fields were assumed to have Gaussian spatial envelopes

$$A_{p,s}(x, y) = \exp\left(-\frac{x^2 + y^2}{w_{p,s}^2}\right), \tag{S31}$$

where $w_p$ and $w_s$ are the 1/e field radii of pump and signal, respectively. The transverse nonlinear polarizations were calculated according to the established spin-selection rules in Eqs. (S22, S23). The Fourier-plane fields were obtained using two-dimensional fast Fourier transforms

$$\tilde{E}_\gamma(k_x, k_y) \propto \mathcal{F}\left\{P_\gamma^{(2)}(x, y)\right\}, \qquad \gamma = SFG, DFG \tag{S32}$$

The corresponding intensity distributions were calculated as

$$I_\gamma(k_x, k_y) = \left|\tilde{E}_\gamma(k_x, k_y)\right|^2. \tag{S33}$$

The spatial-frequency coordinates were converted to transverse wave vectors using $k_x = 2\pi f_x$, $k_y = 2\pi f_y$. The simulations used $\lambda_{SFG} = 344\ nm, \lambda_{DFG} = 603\ nm, w_p = w_s = 0.6\ mm,$ and a default polarization-grating period of $\Lambda = 200\ \mu m$. The SFG and DFG intensity distributions were individually normalized when comparing their momentum-space profiles. Diffraction efficiencies were instead calculated using the unnormalized Fourier intensity and normalized by the total power in the complete Fourier plane

$$\eta_n = \frac{\iint_{\Omega_n} I_{DFG}(k_x, k_y) dk_x dk_y}{\iint I_{DFG}(k_x, k_y) dk_x dk_y}, \tag{S34}$$

where $\Omega_n$ denotes the integration region associated with diffraction order n.

## Note 4 Signal-spin-controlled spectral selection of the nonlinear phase-to-momentum mapping

The main text considers a spatially uniform signal with $\sigma_s = +1$. In this case, the spatially uniform $\sigma^+$ component of the structured pump is mapped onto the SFG pathway, whereas the binary-phase $\sigma^-$ component is mapped onto the DFG pathway. Here, we consider the reversed signal spin $\sigma_s = -1$. According to the established spin-selection rules, reversing the signal spin exchanges the pump-spin components contributing to the SFG and DFG pathways. The nonlinear polarizations become

$$\begin{aligned} \boldsymbol{P}_{SFG}^{(2)}(x, y, \sigma_s = -1) &\propto A_{eff}(x, y) g(x), \\ \boldsymbol{P}_{DFG}^{(2)}(x, y, \sigma_s = -1) &\propto A_{eff}(x, y). \end{aligned} \tag{S35}$$

Consequently, the SFG nonlinear polarization carries the periodic binary $0/\pi$ phase modulation, whereas the DFG nonlinear polarization remains spatially in phase. Their near-field intensity envelopes remain identical in the ideal model

$$\left|P_{SFG}^{(2)}(x, y)\right|^2 = \left|P_{DFG}^{(2)}(x, y)\right|^2 \propto \left|A_{eff}(x, y)\right|^2. \tag{S36}$$

The corresponding Fourier-plane distributions are therefore exchanged. According to Eq. (S28), the SFG field is distributed among the odd diffraction orders, whereas the DFG field is concentrated in the on-axis zeroth order. Reversing the signal spin exchanges the pump-spin components addressed by the SFG and DFG pathways, thereby transferring the binary-phase diffraction pattern from DFG to SFG. As shown in Fig. S2, the binary-phase diffraction pattern is switched from the DFG output to the SFG output, whereas the zeroth-order distribution is switched from the SFG output to the DFG output.

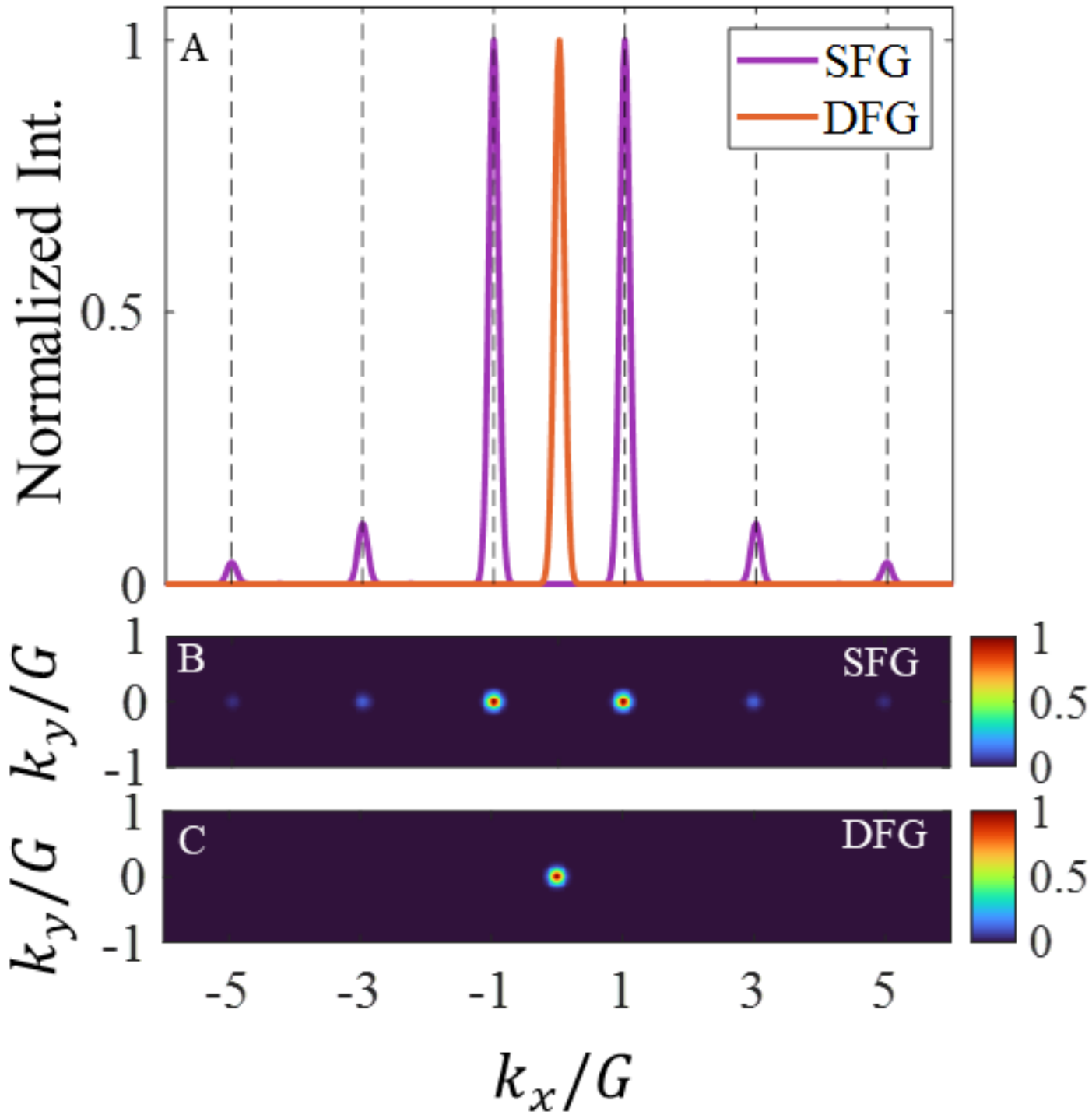

Figure S2 **(A)** Independently normalized Fourier-plane line profiles for a spatially uniform signal with $\sigma_s = -1$. The SFG field is distributed among the odd diffraction orders, whereas the DFG field is concentrated in the zeroth order. Dashed lines indicate the predicted odd diffraction orders $n = \pm1, \pm3, \pm5$. **(B, C)** Corresponding two-dimensional Fourier-plane intensity distributions of the SFG and DFG fields, respectively. The two frequency pathways are normalized independently to compare their momentum-space structures.